\documentclass[10pt,prl,onecolumn,preprint,floats,floatfix,aps,nofootinbib,preprintnumbers]{revtex4-1}
\usepackage{graphicx}
\usepackage{amssymb}
\usepackage{dcolumn}
\usepackage{amsmath}
\usepackage{slashed}
\usepackage{multirow}
\usepackage{natbib}
\usepackage{color}
\usepackage{ulem}
\usepackage{mathrsfs}
\usepackage{stmaryrd}

\usepackage{tikz}
\usetikzlibrary{arrows,decorations.pathmorphing}

\newcommand{\sD}[1]{\begin{tikzpicture}[scale={#1},photon/.style={decorate,decoration={snake,post length=1mm}}]
        \draw (0,4) -- (0.5,2);
        \draw[photon] (0,4) -- (0.5,2);
        \draw node[anchor=east] at (0,4) {\Large $n$};
        \draw (0,0) -- (0.5,2);
        \draw[photon] (0,0) -- (0.5,2);
        \draw node[anchor=east] at (0,0) {\Large $n$};
        \draw (0.5,2) -- (3.5,2);
        \draw (3.5,2) -- (4,4);
        \draw[photon] (3.5,2) -- (4,4);
        \draw node[anchor=west] at (4,4) {\Large $n$};
        \draw (3.5,2) -- (4,0);
        \draw[photon] (3.5,2) -- (4,0);
        \draw node[anchor=west] at (4,0) {\Large $n$};
      \node[circle,fill=darkgray,draw=black,inner sep=0pt,minimum size=0.1cm] at (0.5,2) {};
      \node[circle,fill=darkgray,draw=black,inner sep=0pt,minimum size=0.1cm] at (3.5,2) {};
      \draw node[anchor=south] at (2,2) {\Large $j,\, r$};
\end{tikzpicture}}

\newcommand{\tD}[1]{\begin{tikzpicture}[scale={#1},photon/.style={decorate,decoration={snake,post length=1mm}}]
        \draw (0,4) -- (2,3.5);
        \draw[photon] (0,4) -- (2,3.5);
        \draw node[anchor=east] at (0,4) {\Large $n$};
        \draw (0,0) -- (2,0.5);
        \draw[photon] (0,0) -- (2,0.5);
        \draw node[anchor=east] at (0,0) {\Large $n$};
        \draw (2,0.5) -- (2,3.5);
        \draw (2,3.5) -- (4,4);
        \draw[photon] (2,3.5) -- (4,4);
        \draw node[anchor=west] at (4,4) {\Large $n$};
        \draw (2,0.5) -- (4,0);
        \draw[photon] (2,0.5) -- (4,0);
        \draw node[anchor=west] at (4,0) {\Large $n$};
      \node[circle,fill=darkgray,draw=black,inner sep=0pt,minimum size=0.1cm] at (2,3.5) {};
      \node[circle,fill=darkgray,draw=black,inner sep=0pt,minimum size=0.1cm] at (2,0.5) {};
      \draw node[anchor=east] at (2,2) {\Large $j,\, r$};
\end{tikzpicture}}

\newcommand{\uD}[1]{\begin{tikzpicture}[scale={#1},photon/.style={decorate,decoration={snake,post length=1mm}}]
        \draw (0,4) -- (2,3.5);
        \draw[photon] (0,4) -- (2,3.5);
        \draw node[anchor=east] at (0,4) {\Large $n$};
        \draw (0,0) -- (2,0.5);
        \draw[photon] (0,0) -- (2,0.5);
        \draw node[anchor=east] at (0,0) {\Large $n$};
        \draw (2,0.5) -- (2,3.5);
        \draw (2,3.5) -- (4,0);
        \draw[photon] (2,3.5) -- (4,0);
        \draw node[anchor=west] at (4,0) {\Large $n$};
        \draw (2,0.5) -- (4,4);
        \draw[photon] (2,0.5) -- (4,4);
        \draw node[anchor=west] at (4,4) {\Large $n$};
      \node[circle,fill=darkgray,draw=black,inner sep=0pt,minimum size=0.1cm] at (2,3.5) {};
      \node[circle,fill=darkgray,draw=black,inner sep=0pt,minimum size=0.1cm] at (2,0.5) {};
      \draw node[anchor=east] at (2,2) {\Large $j,\, r$};
\end{tikzpicture}}

\newcommand{\sgD}[1]{\begin{tikzpicture}[scale={#1},photon/.style={decorate,decoration={snake,post length=1mm}}]
        \draw (0,4) -- (2,2);
        \draw[photon] (0,4) -- (2,2);
        \draw node[anchor=east] at (0,4) {\Large $n$};
        \draw (0,0) -- (2,2);
        \draw[photon] (0,0) -- (2,2);
        \draw node[anchor=east] at (0,0) {\Large $n$};
        \draw node[anchor=west] at (4,4) {\Large $n$};
        \draw (2,2) -- (4,0);
        \draw[photon] (2,2) -- (4,0);
        \draw node[anchor=west] at (4,0) {\Large $n$};
        \draw (2,2) -- (4,4);
        \draw[photon] (2,2) -- (4,4);
      \node[circle,fill=darkgray,draw=black,inner sep=0pt,minimum size=0.1cm] at (2,2) {};
      \node[circle,fill=darkgray,draw=black,inner sep=0pt,minimum size=0.1cm] at (2,2) {};
\end{tikzpicture}}

\newcommand{\ahhh}[3]{a_{{#1}\hspace{1 pt}{#2}\hspace{1 pt}{#3}}}

\begin{document}

\title{Sum Rules for Massive Spin-2 Kaluza-Klein\\
Elastic Scattering Amplitudes\footnote{MSUHEP-19-023}}
\author{R. Sekhar Chivukula$^{a,b}$}
\author{Dennis Foren$^{a,b}$}
\author{Kirtimaan A Mohan$^{b}$}
\author{Dipan Sengupta$^{a}$}
\author{Elizabeth H. Simmons$^{a,b}$}
\affiliation{$^{a}$ Department of Physics and Astronomy, 9500 Gilman Drive,
 University of California, San Diego }
 \affiliation{$^{b}$ Department of Physics and Astronomy, 567 Wilson Road, Michigan State University, East Lansing}


\begin{abstract}
{It has recently been shown explicitly that the high-energy scattering amplitude of the longitudinal modes of massive spin-2 Kaluza Klein states in compactified 5-dimensional gravity, which would naively grow like ${\cal O}(s^5)$, grow only like ${\cal O}(s)$. Since the individual contributions to these amplitudes do grow like ${\cal O}(s^5)$, the required cancellations between these individual contributions result from intricate relationships between the masses of these states and their couplings. Here we report the explicit form of these sum-rule relationships which ensure the necessary cancellations for elastic scattering of spin-2 Kaluza Klein states in a Randall-Sundrum model. We consider an Anti-de-Sitter space of arbitrary curvature, including the special case of a toroidal compactification in which the curvature vanishes. The sum rules demonstrate that the cancellations at ${\cal O}(s^5)$ and ${\cal O}(s^4)$ are generic for a compact extra dimension, and arise from the Sturm-Liouville structure of the eigenmode system  in the internal space. Separately, the sum rules at ${\cal O}(s^3)$ and ${\cal O}(s^2)$ illustrate the essential role of the radion mode of the extra-dimensional metric, which is the dynamical mode related to the size of the internal space. }
\end{abstract}
\maketitle

\section{Introduction}
Scattering amplitudes of massive spin-2 particles have been investigated in a variety of contexts throughout the last few decades. The Fierz-Pauli (FP) theory \cite{Fierz:1939ix} has been established as the only way to write down Lorentz invariant spin-2 mass terms without propagating unphysical ghost degrees of freedom. However, the high-energy behavior of the scattering amplitudes in FP theory limits its applicability.

The growth  of the elastic scattering amplitude of massive spin-2 states in the  FP theory (assuming that interactions arise from the weak field expansion of the usual gravitational action) has been computed using a variety of approaches, including  direct calculation \cite{Aubert:2003je}, 
the Stueckelberg formalism \cite{Hinterbichler:2011tt,deRham:2014zqa}, and via deconstruction \cite{ArkaniHamed:2002sp,ArkaniHamed:2003vb}.  For an amplitude which grows like $s^\lambda$ (where $s$ is the center-of-mass energy-squared), the associated energy cutoff scale is $\Lambda_{\lambda}= (m_{g}^{\lambda -1 } M_{Pl})^{1/\lambda}$, where $m_{g}$ is the mass of the spin-2 state and $M_{Pl}$ is the Planck mass (the weak-field coupling).  In FP theory, the scattering amplitudes of the longitudinal polarization helicity states grow like $s^{5}/(m_{g}^{8} M_{Pl}^{2})$, and hence the theory is valid up to an energy scale $\Lambda_{5}\ll M_{Pl}$ (we therefore refer to FP as a $\Lambda_5$ theory). Adding non-linear potential terms (polynomial interactions) can reduce the growth of these amplitudes down to $\mathcal{O} (s^{3})$ \cite{ArkaniHamed:2002sp,ArkaniHamed:2003vb,Schwartz:2003vj,Cheung:2016yqr,Bonifacio:2018vzv,Bonifacio:2018aon}, yielding a $\Lambda_{3}$ theory \cite{deRham:2010kj,deRham:2010ik}.  However, no analog of the Higgs mechanism to further moderate this high-energy growth of scattering amplitudes has been found. Indeed, it has been shown that coupling an arbitrary number of scalars  and vectors to the FP theory does not reduce the growth of tree-level amplitudes below ${\cal O}(s^{3})$ \cite{Bonifacio:2019mgk}.
 
In contrast, in a theory where massive spin-2 particles emerge as part of a tower of Kaluza-Klein (KK) \cite{Kaluza:1921tu,Klein:1926tv} states from compactified extra dimensions, a high-energy growth of order $s^5$ (or even $s^3$) must be  absent. Here the high-energy behavior will be governed by the underlying 5D gravitational theory which must enforce relationships such that all terms  displaying bad high-energy growth (including those which arise from the exchange of other massive spin-2 states in the KK tower) cancel out, leaving the 4D scattering amplitude of the massive spin-2 KK states to  grow only as fast as $\mathcal{O}(s)$. We have recently demonstrated \cite{Chivukula:2019rij} that such cancellations occur both for compactified toroidal theories and for Anti-de-Sitter space ($\rm AdS_{5}$) in the Randall-Sundrum model (RS1) \cite{Randall:1999ee}.  

In this letter, we derive the explicit relationships between the couplings and the masses of the KK states which guarantee the required cancellations occur in the elastic scattering amplitudes. We demonstrate that the relationships needed to ensure the elastic scattering amplitudes grow no faster than ${\cal O}(s^3)$ arise from the Sturm-Liouville form of the KK mode expansion in the internal space, and therefore will naturally generalize to KK theories with internal dimensions with arbitrary internal structure. Interestingly, these relations are closely related to coupling relationships which arise in compactified gauge theories \cite{Csaki:2003dt}.\footnote{Conceivably, this may be due to a relationship between five-dimensional gauge- and gravity-theories \cite{Bern:2008qj}.}  Separately, we show that there are additional sum rules which ensure cancellation at ${\cal O}(s^3)$ and ${\cal O}(s^2)$ in $\rm AdS_5$ for arbitrary internal curvature (including the toroidal case, in which the internal curvature vanishes). These final two sum rules illustrate the essential role of the radion mode of the extra-dimensional metric, which is the field related to the size of the internal space.\footnote{As this work was being submitted, we learned that Bonifacio and Hinterbichler \cite{Bonifacio:2019ioc} had, in parallel and by different methods, derived sum rules for KK scattering for a theory with a Ricci-flat internal space. In contrast, our work focuses on the phenomenologically-relevant case of an internal space with constant negative curvature.}

In the rest of this letter, we walk through the derivation of the relevant coupling relations in a compactified $\rm AdS_5$ space of arbitrary curvature. We lay out the key steps and intermediate milestones of the calculation, as well as displaying the final results; the full details of lengthy expressions in the derivations are reserved to a subsequent publication. We conclude with a discussion of the properties of an effective theory in which the tower of KK modes is truncated, and discuss questions for future investigation.


\section{Metric and Sturm-Liouville Problem}
The starting point of this letter is to analyze the boundary value problem for the gravitational KK modes in the Randall Sundrum 1 (RS1) model. 
The geometry of RS1 \cite{Randall:1999ee} is that of a truncated and orbifolded AdS$_5$  space bounded on either end by UV (Planck) and IR (TeV) branes. Bulk and brane cosmological constant terms  are added to the action to ensure that the effective 4D background remains flat. The interactions (here we consider only the gravitational fields, and do not include matter) come from the 5D Einstein-Hilbert action (plus cosmological constant terms, $S_{\text{CC}}$)
\begin{equation}
S= \frac{2}{\kappa^{2}}\int d^{4} x \,d y  \sqrt{\operatorname{det} G_{MN}}\, R + S_{\text{CC}}~,
\label{eq:action}
\end{equation}
where $x^\mu$ are the coordinates of the four non-compact dimensions; $y\in [-\pi r_c,+\pi r_c]$ is the coordinate on the compact internal space, $G_{MN}$ and $R$ are the five-dimensional metric and Ricci scalar respectively; and the dimensionful coupling $\kappa = 2/M^{-3/2}_5$ is the weak-field expansion parameter fixed by the 5D Planck scale $M_5$. The size of the internal space, $r_c$, is arbitrary -- leading to a massless radion scalar mode as discussed below.

Imposing the orbifold symmetry (identifying points in the internal space under $y \to -y$), the 5D RS1 metric in the Einstein frame can be written \cite{Charmousis:1999rg,Rattazzi:2003ea}
\begin{footnotesize}
\begin{eqnarray}
G_{M N} &=&\left( \begin{array}{cc}{e^{-2(k|y|+\hat{u})}\left(\eta_{\mu \nu}+\kappa \hat{h}_{\mu \nu}\right)} & {0} \\ {0} & {-(1+2 \hat{u})^{2}}\end{array}\right) \quad  \nonumber \\ 
\hat{u} &\equiv &\frac{\kappa \hat{r}}{2 \sqrt{6}} e^{+k\left(2|y|-\pi r_{c}\right)}~,
\label{eq:metric}
\end{eqnarray}
\end{footnotesize}
where $k$ (which has dimensions of mass) is the curvature of the internal AdS$_5$ space.\footnote{The four-dimensional Planck scale is given by $M_{Pl}^{2} = (1-e^{-2kr_{c}\pi})M_{5}^{3}/k$.} Here the 5D fields $\hat{h}_{\mu\nu}(x,y)$ and $\hat{r}(x,y)$ are even functions of $y$, and $\eta_{\mu\nu}$ is the usual (mostly-minus in our convention) Lorentz metric. The limit $k\to 0$ corresponds to a flat internal space, and hence to a compactification on an orbifolded torus. As noted above our results will be true for arbitrary $k$, though physically we require $k < M_5$ in order for the 5D theory to remain a valid effective field theory. 

As usual, we will decompose the 5D fields $\hat{h}_{\mu\nu}$ and $\hat{r}$ via a Kaluza-Klein decomposition, where each is replaced by a sum of harmonic functions (specified below) in the internal space weighted by 4D KK states. The 5D $\hat{h}_{\mu\nu}$ field yields a tower of spin-2 4D states which can be labeled by a `KK' number $n$ equal to the number of nodes of its associated wavefunction on the interval $y \in [0,\pi r_c]$. The spin-2 tower begins includes a massless mode with $n=0$, which is associated with the 4D graviton, as well as an infinite tower of massive spin-2 states with $n>0$ -- in what follows, ``KK mode" will refer specifically to these massive spin-2 states.  Using a suitable gauge \cite{Callin:2004zm}, the 5D field $\hat{r}$ can be made independent of the internal coordinate $y$ -- and hence gives rise to a single 4D (massless) scalar field, the radion. Using this form of $G_{MN}$, and the harmonic expansion defined below, the quadratic terms in the action are diagonal.

We expand the action in terms of the metric in Eq. \ref{eq:metric}, and the 5D field $\hat{h}_{\mu\nu}(x,y)$ in the mode expansion
\begin{equation}
\hat{h}_{\mu\nu}(x,y) = \frac{1}{\sqrt{\pi r_c}}\sum_{n=0}^{\infty} h^{(n)}_{\mu\nu}(x) \psi_n(y)~,
\label{eq:kk-modes}
\end{equation}
where the fields $h^{(n)}_{\mu\nu}(x)$ are the spin-2 massless graviton ($n=0$) and massive KK modes ($n>0$). Diagonalizing the quadratic terms, one finds the internal wavefunctions $\psi_n(y)$ must satisfy \cite{Randall:1999ee}
\begin{equation}
-\frac{d}{d y}\left[e^{-4 k|y|} \frac{d \psi_{n}}{d y}\right]=m_{n}^{2} e^{-2 k|y|} \psi_{n}~,	
\label{eq:sturm-liouville}
\end{equation}
subject to the boundary conditions $\partial_y\psi_n(y=0)=\partial_y\psi_y(y=\pi r_c)\equiv0$. 
The solutions are of the form \cite{Goldberger:1999wh}
\begin{equation}
\psi_{n}(y)=\frac{e^{+2 k|y|}}{N_{n}}\left[J_{2}\left(\frac{m_{n}}{k} e^{+k|y|}\right)+b_{n 2} Y_{2}\left(\frac{m_{n}}{k} e^{+k|y|}\right)\right]~,
\end{equation}
where the $m_n$ are determined by roots of the equation
\begin{equation}
\begin{aligned} 0=( & 2\left.J_{2}\right|_{y=\pi r_{c}}+\frac{m_{n}}{k} e^{+k r_{c} \pi}\left.J_{2}^{\prime}\right|_{y=\pi r_{c}} )\left(2\left.Y_{2}\right|_{y=0}+\frac{m_{n}}{k}\left.Y_{2}^{\prime}\right|_{y=0}\right) \\ &-\left(2\left.Y_{2}\right|_{y=\pi r_{c}}+\frac{m_{n}}{k} e^{+k r_{c} \pi}\left.Y_{2}^{\prime}\right|_{y=\pi r_{c}}\right)\left(2\left.J_{2}\right|_{y=0}+\frac{m_{n}}{k}\left.J_{2}^{\prime}\right|_{y=0}\right)~
\end{aligned}	
\end{equation}
where primes denote the derivative of the corresponding functions.
Eq. \ref{eq:sturm-liouville} (with the boundary conditions) is of Sturm-Liouville form with weight function $e^{-2k|y|}$, and has no degenerate eigenvalues $m^2_n$. Therefore, for appropriate normalization constants $N_n$, the solutions are orthonormal and complete
\begin{eqnarray}
	\frac{1}{\pi r_{c}} \int_{-\pi r_{e}}^{+\pi r_{e}} d y\, e^{-2 k|y|} \psi_m(y)\psi_n(y) & =\delta_{mn}~, \label{eq:orthonormal}\\
\frac{1}{\pi r_c} e^{-2k|y|} \sum_{j} \psi_{j}(y) \psi_{j}(y^{\prime}) & = \delta(y-y^{\prime})	~. \label{eq:completeness}
\end{eqnarray}
Finally, the orthogonality relation (Eq. \ref{eq:orthonormal}), along with the equation (Eq. \ref{eq:sturm-liouville}) and the boundary conditions imply that
\begin{equation}
\frac{1}{\pi r_{c}} \int_{-\pi r_{e}}^{+\pi r_{e}} d y\, e^{-4 k|y|}\left(\partial_{y} \psi_m\right)\left(\partial_{y} \psi_n\right)=m_{n}^{2} \delta_{mn}~,
\end{equation}
insuring that the graviton and KK modes have canonical kinetic energy terms.


Similarly for the 5D radion field, which in a suitable gauge \cite{Callin:2004zm} has no $y$-dependence, we have the expansion
\begin{equation}
\hat{r}(x,y) = \frac{1}{\sqrt{\pi r_c}} r(x)\, \psi_0	
\label{eq:radion}
\end{equation}
where $r(x)$ is the 4D scalar radion field, and $\psi_0$ is the normalized zero-mode internal wavefunction
\begin{equation}
\psi_{0}=\sqrt{\frac{k r_{c} \pi}{1-e^{-2 k r_{c} \pi}}}	~.
\label{eq:zero-mode}
\end{equation}

\section{Coupling Definitions} 
\label{sec:coupdef}
\begin{figure}
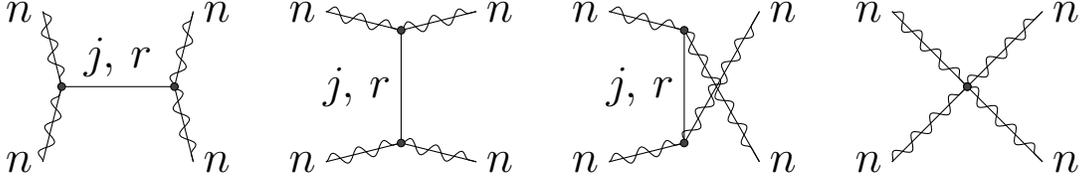

 $\sD{0.5}\hspace{15 pt}\tD{0.5}\hspace{15 pt}\uD{0.5}\hspace{15 pt}\sgD{0.5}$
\caption{Feynman diagrams contributing to $nn \to nn$ level spin-2 KK boson scattering, including $s$-, $t$-, and $u$-channel exchange of KK modes of arbitrary level $j$ or the radion $r$, and four-point contact interactions.}
\label{fig:fd}
\end{figure}
Expanding the action to higher order  in 5D fields and subsequently in terms of the 4D modes, we obtain interactions between the massless graviton, the KK modes, and the radion. The Lorentz form of the interactions are thereby completely determined, but the resulting expressions are lengthy and will be given explicitly in a subsequent publication, as mentioned above.
The coupling strengths of the interactions of the 4D fields are given by overlap integrals of the internal space wavefunctions. There are two classes of these couplings \cite{Davoudiasl:2001uj}, those
 which depend on derivatives  of the internal wavefunctions ({\it i.e.} involving $\partial_y \psi_n$, which we denote\footnote{These $b$-type couplings result, after integrating over the internal space, in non-derivative polynomial interactions of the 4D KK mode fields.} as $b$-type), and those that do not (which we denote as $a$-type).  In general, the 3 and 4 point massive spin-2 KK modes have both $a$-type and $b$-type self-couplings, while couplings of these modes with radions are purely $b$-type.

We will be computing the elastic scattering amplitude $nn \to nn$, where the incoming and outgoing states are both of KK level $n$. At tree-level, this process occurs via a contact interaction, or the exchange of a radion or arbitrary intermediate KK  state $j$ (summed over  $j$ in the complete amplitude). These amplitudes are defined in terms of the following  KK-mode couplings
\begin{eqnarray}
a_{n n j} & =\frac{1}{\pi r_{c}} \int_{-\pi r_{c}}^{\pi r_{c}} d y\, e^{-2 k|y|} \psi_{n}(y) \psi_{n}(y) \psi_{j}(y)~, \label{eq:a3n}	\\
a_{n n n n} & =\frac{1}{\pi r_{c}} \int_{-\pi r_{c}}^{\pi r_c} d y\, e^{-2 k|y|} \psi_{n}(y) \psi_{n}(y) \psi_{n}(y) \psi_{n}(y)~, \label{eq:a4n}\\
b_{n n j} & =\frac{r_{c}}{\pi} \int_{-\pi r_{c}}^{\pi r_{c}} d y\, e^{-4 k|y|}\left(\partial_{y} \psi_{n}(y) \partial_{y} \psi_{n}(y)\right) \psi_{j}(y)~,\label{eq:b3n1}  \\
b_{n j n} & =\frac{r_{c}}{\pi} \int_{-\pi r_{c}}^{\pi r_{c}} d y\, e^{-4 k|y|}\left(\partial_{y} \psi_{n}(y) \partial_{y} \psi_{j}(y)\right) \psi_{n}(y)~, \label{eq:b3n2} \\
b_{n n n n} & =\frac{r_{c}}{\pi} \int_{-\pi r_{c}}^{\pi r_{c}} d y\, e^{-4 k|y|}\left(\partial_{y} \psi_{n}(y) \partial_{y} \psi_{n}(y)\right) \psi_{n}(y) \psi_{n}(y)~,
\label{eq:b4n}
\end{eqnarray}
and the coupling of a radion to two KK-modes
\begin{equation}
b_{n n r}=\frac{r_{c}}{\pi}\, e^{-\pi k r_c} \int_{-\pi r_{c}}^{\pi r_{c}} d y\, e^{-2 k|y|}\left(\partial_{y} \psi_{n}(y) \partial_{y} \psi_{n}(y)\right) \psi_{0}~.	
\label{eq:couprad}
\end{equation}
As suggested by the notation and as shown in Fig. \ref{fig:fd}, the couplings $a_{nnj}$ and $b_{nnj}$ mediate tree-level $s$-, $t$-, and $u$-channel diagrams with intermediate states of level $j$, while $b_{nnr}$ does the same for radion intermediate states, and $a_{nnnn}$ and $b_{nnnn}$ represent four-point ``contact" interactions between four KK-modes of level $n$.

From the definitions of the coupling constants, and using Eq. \ref{eq:sturm-liouville} and the boundary conditions, integration-by-parts yields the following relations
\begin{eqnarray}
b_{nnj} & = \left(m^2_n - \frac{1}{2} m^2_j\right)r^2_c\,a_{nnj}~, \label{eq:rule1}\\
b_{njn} & = \frac{1}{2} m^2_j r^2_c\, a_{nnj}~, \label{eq:rule2} \\
b_{nnnn} & = \frac{1}{3} m^2_n r^2_c\, a_{nnnn} \label{eq:rule3}~.
\end{eqnarray}
\section{Scattering Amplitudes and Sum-Rules}
\label{sec:sumrules}
Using these couplings and the interactions derived from the action, we compute
the Feynman amplitude of the longitudinal (helicity-0) states for   $nn \to nn$ KK-mode scattering and then perform a Laurent expansion to isolate the contributions with differing rates of growth for a center-of-mass energy-squared  $s$ and scattering angle $\theta$,
\begin{equation}
\mathcal{M}(s,\cos\theta)\equiv \sum_{k\le 5} \mathcal{M}^{(k)}(\cos\theta) \cdot s^{k}~. \label{eq:Laurent}
\end{equation}
In the following, we examine the conditions on the couplings which ensure that ${\mathcal M}^{(k)}$ vanishes for $k\in\{2,3,4,5\}$.

At ${\cal O}(s^5)$, applying Eqs. \ref{eq:rule1} and \ref{eq:rule2}, we find
\begin{equation}
{\mathcal M}^{(5)}(\cos\theta)= -\,\frac{\kappa^2}{\pi r_c}\frac{(7+\cos2\theta)\sin^2\theta }{2304  m^8_n}\,\cdot\, \left(a_{nnnn} -\sum_j a^2_{nnj}\right)~.
\end{equation}
Using completeness, Eq. \ref{eq:completeness}, we find that
\begin{equation}
a_{nnnn} = \sum_j a^2_{nnj}~,
\label{eq:sumrule1}
\end{equation}
and hence ${\mathcal M}^{(5)}$ vanishes identically. 

We next  look at the ${\cal O}(s^4)$ piece, where we find, 
\begin{equation}
{\mathcal M}^{(4)}(\cos\theta)= \frac{\kappa^2}{\pi r_c}\frac{(7+\cos2\theta)^2}{27648  m^8_n}\, \cdot \, \left(4 m^2_n a_{nnnn}  - 3 \sum_j m^2_j a^2_{nnj} \right)~.
\end{equation}
Using the Sturm-Liouville equation, integrating by parts twice, and using the boundary conditions we find
\begin{equation}
m^2_j a_{nnj} =	-\frac{2}{r^2_c} b_{nnj}+2m^2_n a_{nnj}~.
\end{equation}
Hence
\begin{align}
\sum_j m^2_j a^2_{nnj} & = \sum_j a_{nnj} \left(-\frac{2}{r^2_c} b_{nnj}+2m^2_n a_{nnj}\right)\\
& = -\frac{2}{r^2_c} b_{nnnn}+2m^2_n a_{nnnn}\\
& = \frac{4}{3}m^2_n a_{nnnn}~.
\end{align}
Here the second line follows from completeness and the last line from Eq. \ref{eq:rule3}. Consequently,
we find a second sum rule
\begin{equation}
m^2_{n} a_{nnnn} = \frac{3}{4} \sum_j m^2_j a^2_{nnj}~,	
\label{eq:sumrule2}
\end{equation}
that ensures that ${\mathcal M}^{(4)}$ also vanishes identically.

The sum rules above (Eqs. \ref{eq:sumrule1} and \ref{eq:sumrule2}) follow directly from the Sturm-Liouville structure of the harmonic expansion for the spin-2 KK fields.  Therefore, these rules will apply to internal spaces of arbitrary warping and size. It is also notable that, having applied Eqns. \ref{eq:rule1} - \ref{eq:rule3} (that is, expressing the sum rules purely in terms of $a$-type couplings), these relations are {\it identical} to coupling relationships which arise in compactified gauge theories \cite{Csaki:2003dt}.
 
 The situation changes at ${\cal O}(s^3)$, however, where we find
\begin{equation}
{\mathcal M}^{(3)}(\cos\theta)= \frac{\kappa^2}{\pi r_c}\frac{\sin^2\theta}{3456 m^8_n}\, \cdot \, 
\left( -108\frac{b^2_{nnr}}{r^4_c} + 12 m^4_n a^2_{nn0}-16 m^4_n a_{nnnn}+15 \sum_j m^4_j a^2_{nnj}\right)~.
\end{equation}
We find explicitly that the radion begins to contribute at this order, as expected from \cite{Schwartz:2003vj}. The
vanishing of this contribution enforces the sum rule
\begin{equation}
\frac{b^2_{nnr}}{r^4_c}=\frac{1}{9} m^4_n a^2_{nn0} -\frac{4}{27} m^4_n a_{nnnn} + \frac{5}{36} \sum_j m^4_j a^2_{nnj}~,
\label{eq:sumrule3}
\end{equation}
which relates the radion coupling to the spin-2 KK-mode self-couplings, and therefore relies on the structure of the action and the properties of the radion. While we have not found a general analytic derivation of this expression, we have numerically verified that this sum rule is satisfied for many values of $kr_c$ -- including intermediate values such as $kr_c \simeq 2$ which are far from both the decoupling limit ($kr_c \gg 1$, which corresponds to $M_{Pl} \to \infty$) and from $k r_c =0$ corresponding to an orbifolded torus. 

For terms at ${\cal O}(s^2)$ we find
\begin{align}
	{\mathcal M}^{(2)}(\cos\theta) &= -\, \frac{\kappa^2}{\pi r_c}  \frac{(7+\cos2\theta)}{5184 m^8_n}\, \cdot \,  \left( -108 \frac{b^2_{nnr} m^2_n}{r^4_c} + 12 m^6_n a^2_{nn0}  \right. \nonumber \\
	& \left.\phantom{\frac{b^2_{nnr} m^2_n}{r^4_c}} + 16 m^6_na_{nnnn} - 15 m^2_n \sum_j m^4_j a^2_{nnj}+6 \sum_j m^6_j a^2_{nnj} \right)~.
\end{align}
The vanishing of this contribution yields a final sum rule
\begin{equation}
\frac{b^2_{nnr} m^2_n}{r^4_c}=\frac{1}{9} m^6_n a^2_{nn0} +\frac{4}{27} m^6_n a_{nnnn} - \frac{5}{36} m^2_n \sum_j m^4_j a^2_{nnj}+\frac{1}{18}\sum_j m^6_j a^2_{nnj}~.
\label{eq:sumrule4}
\end{equation}
Remarkably, the same combination of radion and massless graviton coupling appears in Eqs. \ref{eq:sumrule3} and \ref{eq:sumrule4}, 
and therefore inserting the order $s^{3}$ relationship into the equation above, we also obtain the following relation, 
\begin{equation}
   \sum_{j=0}^{+\infty}\hspace{3 pt} m_j^6 \hspace{3 pt} \ahhh{n}{n}{j}^{2}  =  5 m^2_n \sum_{j=0}^{+\infty}\hspace{3 pt} m_j^4 \hspace{3 pt} \ahhh{n}{n}{j}^{2}~-\frac{16}{3}m^6_na_{nnnn},   \label{eq:sumrule5}
\end{equation}
which depends only on the spin-2 modes.
In the case of RS1 (with exponential warping), Eq. \ref{eq:sumrule5} can be derived analytically from the form of the wavefunctions.

Finally, the non-zero amplitude ${\mathcal M}^{(1)}$ is given by
\begin{align}
{\mathcal M}^{(1)}(\cos\theta) & = \frac{\kappa^2}{\pi r_c}  \frac{(7+\cos2\theta)^2 \csc^2\theta }{34560 m^8_n}\, \cdot \, \left( -1296 \frac{b^2_{nnr} m^4_n}{r^4_c} +144 m^8_n a^2_{nn0}\right. \nonumber \\
&\left.\phantom{\frac{b^2_{nnr} m^2_n}{r^4_c}} +28 m^8_n a_{nnnn}+15 \sum_j m^8_j a^2_{nnj}\right)~.
\label{eq:soneamplitude}
\end{align}
Numerically evaluated, the $\mathcal{O}(s)$, term grows like $\sim s/\Lambda_{\pi}^{2} $ \cite{Chivukula:2019rij}, where $\Lambda_{\pi}\equiv M_{Pl}e^{-k r_{c}\pi }$, and therefore we estimate $\Lambda_{\pi}$ to be the cut off scale of the theory, as inferred from the AdS/CFT \cite{Maldacena:1997re,Witten:1998qj,Aharony:1999ti,ArkaniHamed:2000ds} correspondence. In the limit when $k\to 0$, i.e, for a flat extra dimension, the amplitude grows as $\sim s/M_{Pl}^{2} $.

\section{Discussion}

The above set of sum rules (Eqs. \ref{eq:sumrule1}, \ref{eq:sumrule2}, \ref{eq:sumrule3} and \ref{eq:sumrule4}) describe the necessary coupling and mass relations required to cancel all helicity-0 $nn \to nn$ elastic matrix elements at every order in $s^{k}$, starting from 
$\mathcal{O}(s^{5})$  down to $\mathcal{O}(s)$.\footnote{We note that our sum-rules are equivalent to the ``bottom-up" sum rules proposed in \cite{Bonifacio:2019ioc}, when applied to AdS$_5$.} Although we have described here only the scattering of the longitudinal helicity  states, we have also verified that these relations are sufficient to ensure that the elastic scattering of all helicity states cancels down to $\mathcal{O}(s)$ as well.\footnote{Details of transverse polarization scattering will be provided in a subsequent publication.} 

The ${\cal O}(s)$ high-energy behavior \cite{Chivukula:2019rij} of the massive spin-2 scattering  amplitudes opens up new parameter space to explore in phenomenologically motivated models of particle physics and dark matter (for example, see \cite{Chivukula:2017fth} and references therein). The sum rules imply that an accurate calculation of the relevant cross sections in such theories must appropriately model the contributions from the radion and higher-mass KK states -- ``simplified models" including only a single massive spin-2 state will have only limited validity.
 
 These sum rules derived in the context of an infinite tower of KK states also allow us to consider the behavior of elastic scattering amplitudes for a KK theory truncated at level $N_{max}$. Since Eqs.  \ref{eq:sumrule1} and  \ref{eq:sumrule2} rely on the completeness  relations. these will not be fully satisfied in the truncated theory, and therefore there are finite  $\mathcal{O} (s^{5})$ contributions in the amplitudes. For nonzero curvature,  these contributions are suppressed as $\frac{1}{N_{max}^{2k +1 }}$ \cite{Chivukula:2019rij} for a matrix element  ${\mathcal M}^{(k)}$. Such corrections are therefore small and vanish in the limit $N_{max}\to \infty$. In the limit when $k\to 0$, {\it i.e.} for a flat (toroidal) compactification, the intermediate propagators can only be either level $0$ (the massless graviton) or level $2n$, and the above sum rules only involve two intermediate states due to KK number conservation. In such a scenario,  if the theory is truncated at $N_{max}< 2n$, the scattering amplitudes grow as $\mathcal{O}(s^{5})$, i.e the theory becomes strongly coupled at $\Lambda_{5}$, contrary to the previous speculation of  $\Lambda_{3}$ \cite{Schwartz:2003vj}. 

A number of additional questions remain to be explored. For example, Eqs. \ref{eq:sumrule3} and \ref{eq:sumrule4} involve both the radion and the KK-mode couplings, and must follow from the form of the action and the properties of the radion in a way yet to be uncovered. Any realistic model of a compactified extra dimensional theory requires a mass for the radion, and a stabilization mechanism for the geometry. However, putting in a mass for the radion, assuming no changes in the radion couplings, would reintroduce amplitudes which grow like ${\cal O}(s^2)$ \cite{Schwartz:2003vj}. Therefore, a realistic model with a stablized geometry, (e.g. \cite{Goldberger:1999uk}) must give rise to additional contributions to the sum rule in Eq. \ref{eq:sumrule4}. Finally, phenomenologically-relevant models also include matter fields (either localized to a brane, or in the bulk). The contributions from these fields will contribute at order ${\cal O}(s^3)$ and ${\cal O}(s^2)$ and will therefore modify the sum rules presented here. Details of the calculations above and further consideration of these questions will be given in a forthcoming publication.

This material is based upon work supported by the National Science Foundation under Grant No. PHY-1915147 .

\end{document}